# Heavy-fermion superconductivity in $CeCoIn_5$ at 2.3 K

C. Petrovic[1], P.G. Pagliuso[2], M.F. Hundley[2], R. Movshovich[2], J.L. Sarrao[2],
J.D. Thompson[2], Z. Fisk[1,2], and P. Monthoux[3]

[1]National High Magnetic Field Laboratory,
Florida State University, Tallahassee, FL 32306 USA

[2]Condensed Matter and Thermal Physics,
Los Alamos National Laboratory, Los Alamos, NM 87545 USA

[3]Cavendish Laboratory, University of Cambridge, Cambridge CB3 OHE, UK

Abstract

We report the observation of heavy-fermion superconducitivity in $CeCoIn_5$ at $T_c$=2.3 K. When compared to the pressure-induced $T_c$ of its cubic relative $CeIn_3$ ($T_c$~200 mK), the $T_c$ of $CeCoIn_5$ is remarkably high. We suggest that this difference may arise from magnetically mediated superconductivity in the layered crystal structure of $CeCoIn_5$.

Superconductivity is distinct in the correlation often evident between structure and properties: certain crystal structures or substructures favor superconductivity.[1] In particular, what underlies this relationship in the high-$T_c$ cuprates and heavy-Fermion materials, which border so closely on magnetically ordered phases, is of essential interest both fundamentally and in the search for new superconducting materials.[2,3] For example, fully half of the known heavy-Fermion superconductors crystallize in the tetragonal $ThCr_2Si_2$ structure, which is also the structure type of the $La_2CuO_4$ family of high-$T_c$ superconductors.[4] In the cuprates, there is no consensus on the origin of superconductivity, but their quasi-2D structure and proximity to magnetic order have been shown to be particularly favorable for an unconventional form of superconductivity in which a pairwise attractive interaction among quasiparticles is mediated by magnetic correlations.[5] Here, we report the discovery of a possible heavy-Fermion analogue of the cuprates, a new layered superconductor $CeCoIn_5$, with the highest known ambient-pressure superconducting transition temperature $T_c$ in the class of heavy-Fermion materials.

Heavy Fermion superconductors are materials in which superconductivity emerges out of a normal state with strong electronic correlations. The presence of an appropriate magnetic ion – in this case Ce – enhances the effective mass m* of conduction electrons by several orders of magnitude.[6] In the more than twenty years since the first heavy Fermion superconductor was discovered ($CeCu_2Si_2$),[7] only one other Ce-based material has been found that unambiguously shows superconductivity at ambient pressure: $CeIrIn_5$.[8] Both of these materials exhibit rather complex phenomena and/or metallurgy, making their study challenging. The ground state of $CeCu_2Si_2$ can be either antiferromagnetic or superconducting depending on very small changes in unit-cell volume or composition;[9] CeIrIn$_5$ shows zero resistivity near 1 K but does not produce a thermodynamic signature of superconductivity until 0.4 K.[8] Although the complexity of these materials is a direct manifestation of the richness of their physics, our fundamental understanding of the nature of heavy-Fermion superconductivity would profit substantially from the existence of a simpler example. CeCoIn$_5$ appears to be such a material.

CeCoIn$_5$ forms in the tetragonal HoCoGa$_5$ crystal structure with lattice constants a=4.62 Å and c=7.56 Å.[10] The structure is built up of alternating layers of CeIn$_3$ (a heavy Fermion antiferromagnet in which superconductivity can be induced with applied pressure[3]) and 'CoIn$_2$,'[11] and is analagous to its isostructural relatives CeRhIn$_5$[12] and CeIrIn$_5$.[8] Single crystals of CeCoIn$_5$ were synthesized from an In flux by combining stoichiometric amounts of Ce and Co with excess In in an alumina crucible and encapsulating the crucible in an evacuated quartz ampoule. Because of the deep eutectic formed between Ce and Co and the strong phase stability of CeIn$_3$, growth of CeCoIn$_5$ appears to be optimized in dilute (3 at% Ce) melts with a two-stage cooling process – an initial rapid cooling from 1150 C, where the molten material is homogenized, to 750 C and then a slower cool to 450 C, at which point the excess flux is removed with a centrifuge. The resultant crystals are well-separated, faceted platelets with characteristic dimensions 3 mm x 3 mm x 0.1 mm.

Figure 1 shows the magnetic susceptibility $\chi$ and electrical resistivity $\rho$ of CeCoIn$_5$. The magnetic susceptibility is anisotropic, with $\chi$ larger for a magnetic field applied along the tetragonal c-axis. The rapid upturn at low temperature for $\chi \parallel$ c is intrinsic and is a common feature of the CeMIn$_5$ materials (see Figure 3 below). At high temperatures (T>200), $\chi^{-1}$ is linear in temperature and a paramagnetic Weiss temperature of –83 K (-54 K) is found when the magnetic field is applied parallel (perpendicular) to the c-axis. The effective moment obtained from a polycrystalline average of these data is 2.59 $\mu_B$, consistent with the free-ion expectation for $Ce^{+3}$ (2.54 $\mu_B$). The resistivity of $CeCoIn_5$ is typical of heavy-Fermion materials: weakly temperature dependent above a characteristic temperature, here ~ 30 K, followed by a rapid decrease at lower temperature. This behavior is generally attributed to a crossover from strong, incoherent scattering of electrons at high temperature to the development of strongly correlated Bloch states at low temperature. Below ~ 20 K the resistivity of $CeCoIn_5$ is nearly linear in temperature, a functional dependence found commonly in magnetically mediated superconductors and frequently associated with proximity to a quantum critical point.[3] The low value of $\rho$ at 2.5 K (3 $\mu\Omega$cm) indicates minimal defect scattering. The inset of Figure 1 reveals clear evidence for superconductivity: zero resistance and full-shielding diamagnetism is achieved at 2.3 K.

The specific heat divided by temperature for $CeCoIn_5$ is shown in Figure 2. For T>2.5 K, the large value of C/T=290 mJ/molK² indicates substantial mass renormalization (C/T = $\gamma \propto m^*$). At $T_c$ a surprisingly large jump in heat capacity is observed. The inferred value of $\Delta C/\gamma T_c$=4.5 suggests extremely strong coupling. (The expectation for a weak-coupling superconductor is $\Delta C/\gamma T_c$=1.43.[13]) However, application of a magnetic field of 50 kOe along the c-axis suppresses superconductivity and reveals that entropy is conserved between the normal and superconducting states by 2.3 K because of the continued increase of C/T with decreasing temperature. Because there is presently no theory of superconductivity that directly accounts for a temperature dependent $\gamma$ below $T_c$, this effect calls into question simple estimates of whether $CeCoIn_5$ is a weak- or strong-coupling superconductor. The enhanced normal-state $\gamma$ (~ 1J/molK$^2$) in 50 kOe is also clear evidence that CeCoIn$_5$ is indeed a heavy Fermion material. The temperature dependence of C below T$_c$ is non-exponential, suggesting the presence of nodes in the superconducting gap. (Power law fits of the form C/T = $\gamma_0$ + aT$^x$ at low temperature yield $\gamma_0$ ~ 0.04 J/molK$^2$ and x~1, with the precise value of x depending on the range of temperatures fitted.) Both the size of the jump at T$_c$ and the field-induced normal state temperature dependence of C/T at low temperature (below the zero-field T$_c$) are reminiscent of UBe$_{13}$, another heavy Fermion superconductor.[14]

The inset of Figure 3 presents an H-T phase diagram for CeCoIn$_5$. For both heat capacity and resistivity measurements, the magnetic field is applied parallel to the c-axis. Because of the large $\Delta$C at the superconducting transition, one might suspect that the observed superconductivity is parasitic to a magnetic phase. However, the onset of zero resistance tracks the heat capacity feature identically as a function of field, and the observed transitions remain sharp at the highest fields, arguing against a co-existing magnetic phase. Additional resistivity measurements with H $\perp$ c reveal an upper-critical-field anisotropy of at least a factor of two.

CeCoIn$_5$ appears to be a comparatively simple realization of heavy-Fermion superconductivity with T$_c$=2.3 K. But, why should T$_c$ in this material be over a factor of two higher than that of CeCu$_2$Si$_2$ or CeIrIn$_5$ and ten times that of CeIn$_3$? Furthermore, how does one rationalize the existence of three heavy-Fermion superconductors (CeRhIn$_5$ under pressure, CeIrIn$_5$, and CeCoIn$_5$) in the same structure type?

Recent model calculations[15] for magnetically mediated superconductivity suggest a possible answer. While the origin of superconductivity in heavy-Fermion materials is unknown, there is growing evidence that it is magnetically mediated[3,4]. This seems to be the case as well in the CeMIn$_5$ family. The appearance of superconductivity in CeRhIn$_5$ as magnetism is suppressed by pressure certainly suggests that spin fluctuations are important. Similarly, the power-law dependence of specific heat in CeCoIn$_5$ (Fig. 2) and CeIrIn$_5$ is a hallmark of an unconventional gap function with nodal structure.

For magnetically mediated superconductivity on the border of antiferromagnetism, the pairing interaction is long-ranged and oscillatory in space. In some regions it is attractive but in others repulsive, and therefore tends to cancel on average. Although this might at first sight rule out the possibility of high $T_c$s resulting from such an interaction, it is possible to construct a Cooper-pair wavefunction that only has a significant probability in regions where the oscillatory potential is attractive. The repulsive regions of the potential are neutralized and calculations show that, within Eliashberg theory, magnetic pairing can be strong enough to explain high $T_c$s as high as those observed in the cuprates.[5,16]

In these calculations, quasi-2D tetragonal or orthorhombic crystal structures with a nearly half-filled band, a large Fermi surface, and nearly commensurate antiferromagnetic correlations are essential for producing high $T_c$s. The choice of a Cooper-pair wavefunction with nodes along the diagonals x=±y and opposite phase along the x and y directions almost perfectly matches the oscillations of the pairing potential. The situation is quite different for 3D cubic crystal structures. In this case one needs to match the oscillations of the pairing potential not only in the x-y plane but also in the x-z and y-z planes. By analogy with the quasi-2D case, to match the oscillations in the x-y plane, the Cooper-pair wavefunction must change sign upon a 90 degree rotation about the z-axis. Suppose that the Cooper pair wavefunction is positive along the x-axis and negative along the y-axis. In order to achieve the same effect in the x-z plane, the Cooper-pair wavefunction must be negative along the z-axis. But the need to match the oscillations of the pairing potential in the y-z plane as well requires the Cooper-pair wavefunction to be positive along the z-axis. Thus, it is not possible to avoid cancellations resulting from the oscillatory nature of the pairing potential and lower values of $T_c$ are expected.

There is another effect that favors magnetically mediated superconductivity in quasi-2D over 3D systems. When the magnetism is quasi-2D, there is a line of soft magnetic fluctuations in the Brillouin zone, whereas there is only a point when the magnetism is fully 3D. This leads to an enhanced coupling to the magnetic fluctuations in quasi-2D systems relative to their 3D counterparts. This not only favors higher transition temperatures in quasi-2D materials but also greater robustness of pairing to material imperfections and other competing interaction channels such as electron-phonon scattering.

Although the intrinsic electronic anisotropy in the $CeMIn_5$ materials remains to be determined definitively, structural layering should produce electronic and magnetic states that are less 3D than than in cubic $CeIn_3$. Preliminary de Haas-van Alphen (dHvA) and nuclear quadrupole resonance measurements are beginning to confirm this intuition.[17,18] The anisotropy of $H_{c2}$ in $CeCoIn_5$ provides evidence for electronic anisotropy of at least a factor of four (simple models estimate $m^*_a/m^*_c \sim (H_{c2}^a/H_{c2}^c)^2$), and dHvA measurements reveal that the Fermi surface of $CeCoIn_5$ is substantially more two-dimensional than that of $CeIn_3$ and to a lesser extent than that of $CeRhIn_5$ or $CeIrIn_5$.[19]

The relevant temperature scale for magnetically mediated superconductivity is $T_{sf}$, the characteristic spin fluctuation temperature. An estimate of $T_{sf}$ comes from $\gamma$, which typically is assumed to scale as $1/T_{sf}$. Among the $CeMIn_5$ family, $\gamma$ at $T_c$ increases monotonically from 290 to 400 to 750 mJ/molK$^2$ in the sequence M=Co, Rh, Ir, and $T_c$ decreases as expected in the same sequence. An apparently related trend is seen in Figure 4, in which we show anisotropic magnetic susceptibility data for the $CeMIn_5$ (M = Rh,Ir,Co) materials. These data suggest that members of this family are related and that some parameter relevant to superconductivity is being optimized as the c-axis susceptibility evolves from M=Rh (antiferromagnetic with largest $\chi_c$) to M=Co (superconducting with highest $T_c$). Thus one seems to have found at least a local optimization for magnetically mediated superconductivity in $CeCoIn_5$ intrinsically, whereas tuning with pressure is required for $CeRhIn_5$ to reach a comparable $T_c$. Measurements of $\chi \parallel c$ for $CeRhIn_5$ as a function of pressure would be valuable in further establishing this point.[12]

From a more global perspective, $CeCoIn_5$ also appears to display a rather optimized version of magnetically mediated superconductivity. The same Eliashberg treatment of magnetically mediated superconductivity discussed above predict that $T_c$s of order 20-30% of $T_{sf}$ are the best one can hope to achieve.[15] This prediction is consistent with experience in the cuprates where $T_c$s of order 100 K are achieved relative to $T_{sf}$s of order 700 K (i.e., $T_c/T_{sf}$~0.14).[5] Based on both the value of heat capacity $\gamma$ and the position in temperature of the maximum in resistivity, one can deduce $T_{sf}$~10K for the

CeMIn$_5$ materials, so that in CeCoIn$_5$ one achieves T$_c$/T$_{sf}$~0.2, a value comparable to that achieved in the cuprates and quite close to the maximum expected value. In contrast, cubic CeIn$_3$ achieves a T$_c$ of only 0.2 K despite a T$_{sf}$ in the range 50 – 100 K (so that T$_c$/T$_{sf}$~0.002).[3,20]

Finally, we conclude on a speculative note. Some instability generally limits how high T$_c$ can be in a structurally related family (e.g., the series Nb$_3$Sn, Nb$_3$Ge, Nb$_3$Si).[21] That the properties of CeCoIn$_5$ are relatively simple perhaps suggests that even higher T$_c$ members of this family remain to be found. In Ce-based materials, the spin-fluctuation temperature tends to be low because of the narrow f-bands. In order to increase T$_{sf}$ and hence T$_c$, one would like to move to d-band metals, because they typically have wider bandwidths and hence much larger T$_{sf}$'s. A d-electron analogue of CeCoIn$_5$ with a much higher T$_c$ may be awaiting discovery.

We thank A.V. Balatsky, L.P. Gor'kov, M.J Graf, G.G. Lonzarich, and D. Pines for valuable conversations. This work was stimulated in part by a workshop sponsored by the Institute for Complex Adaptive Matter, supported by the NSF and FAPESP, and performed under the auspices of the U.S. Dept. of Energy.

Figure Captions

Figure 1. Magnetic susceptibility and electrical resistivity of $CeCoIn_5$. Susceptibility is measured in a 1-kOe field applied parallel (circles) or perpendicular (squares) to the c axis of $CeCoIn_5$ using a SQUID magnetometer. The inset shows zero-field-cooled magnetic susceptibility (circles) as a fraction of $1/4\pi$ measured in 10 Oe and resisitivity (triangles) in the vicinity of the superconducting transition.

Figure 2. Specific heat divided by temperature versus temperature for $CeCoIn_5$. For both the zero-field (open squares) and 50-kOe (solid circles) data, a nuclear Schottky contribution, due to the large nuclear quadrupole moment of In, has been subtracted. The inset shows the entropy recovered as a function of temperature in the superconducting (open squares) and field-induced normal (solid circles) states.

Figure 3. Magnetic field-temperature phase diagram of $CeCoIn_5$. The filled diamonds correspond to specific heat measurements, and the open diamonds were determined by electrical resistivity. In both cases magnetic field was applied along the c axis of the sample.

Figure 4. Anisotropic magnetic susceptibility of $CeRhIn_5$ (circles), $CeIrIn_5$ (squares), and $CeCoIn_5$ (triangles) measured in 1 kOe. The susceptibility perpendicular to the c-axis (open symbols) is independent of M, whereas the susceptibility parallel to the c-axis (filled symbols) decreases and changes in character as ambient-pressure $T_c$ increases.

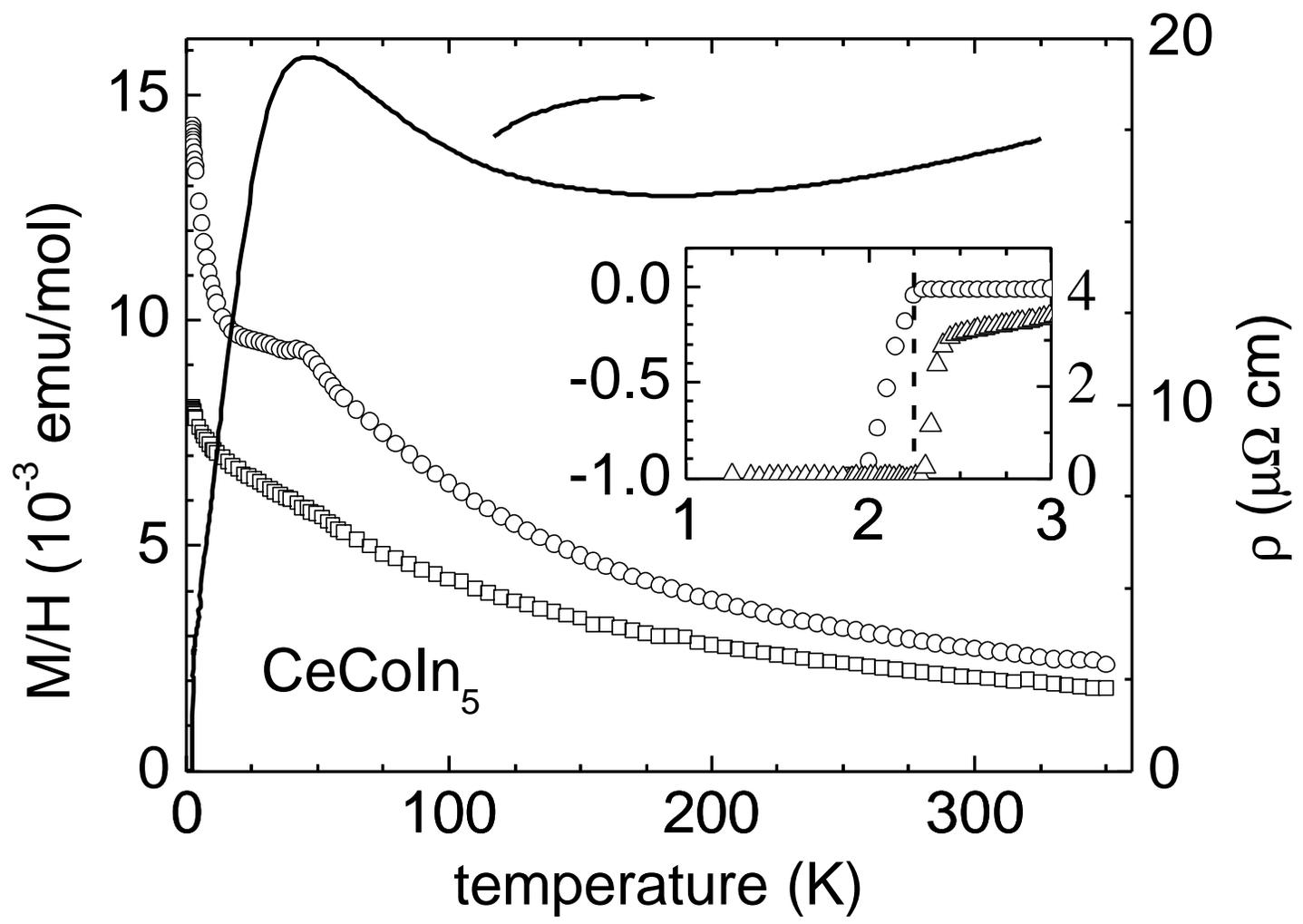

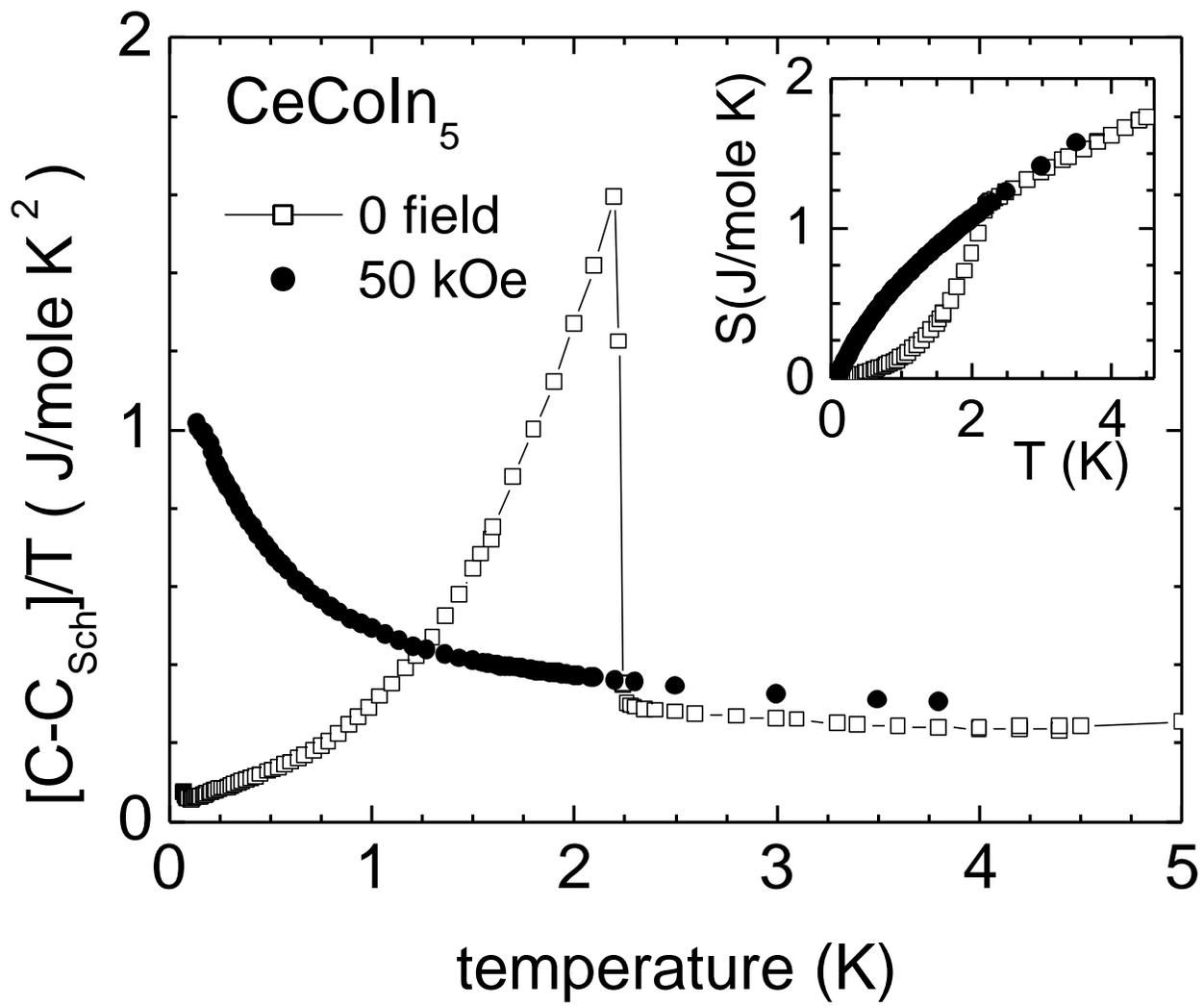

Fig. 2 Petrovic, *et al.*

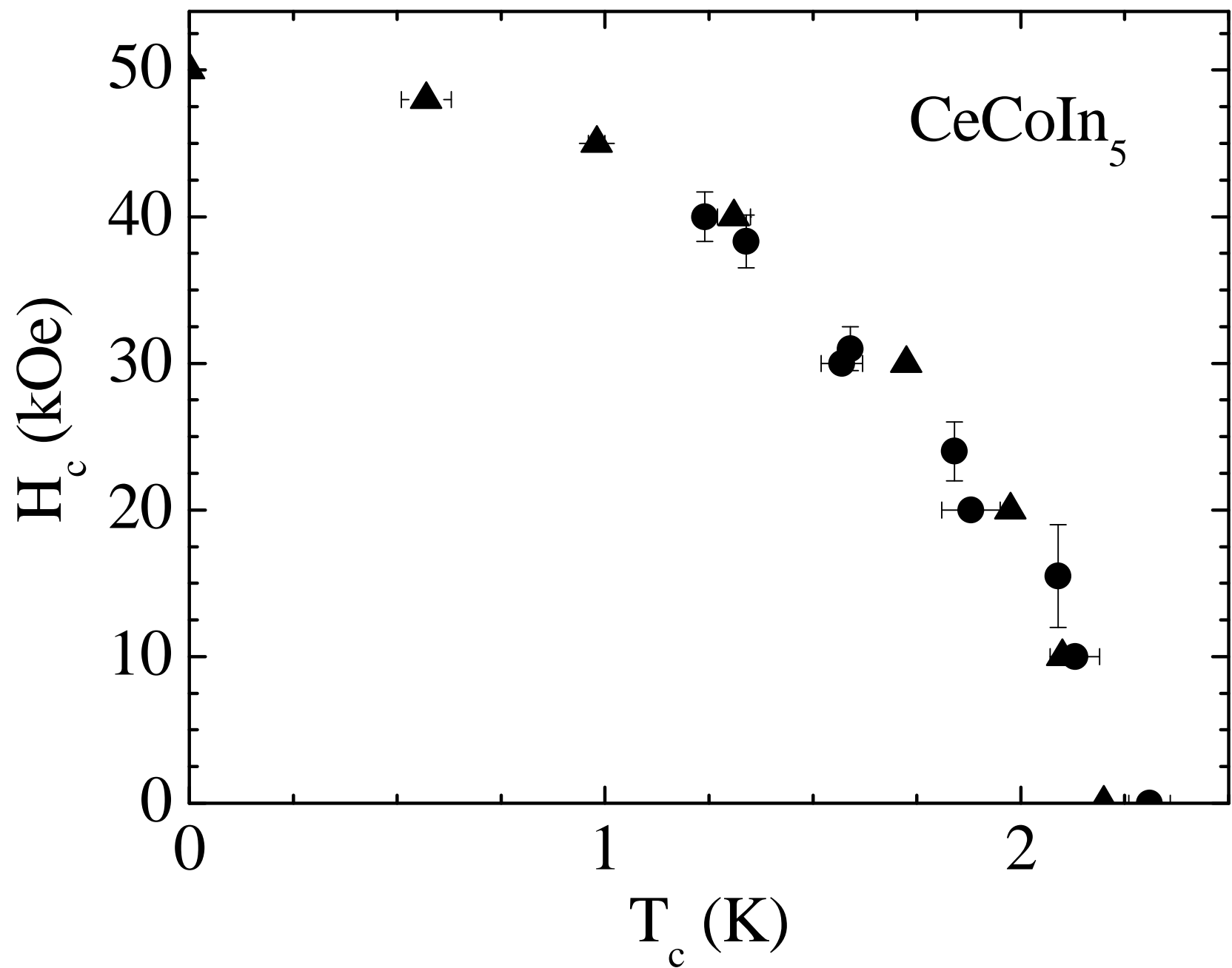

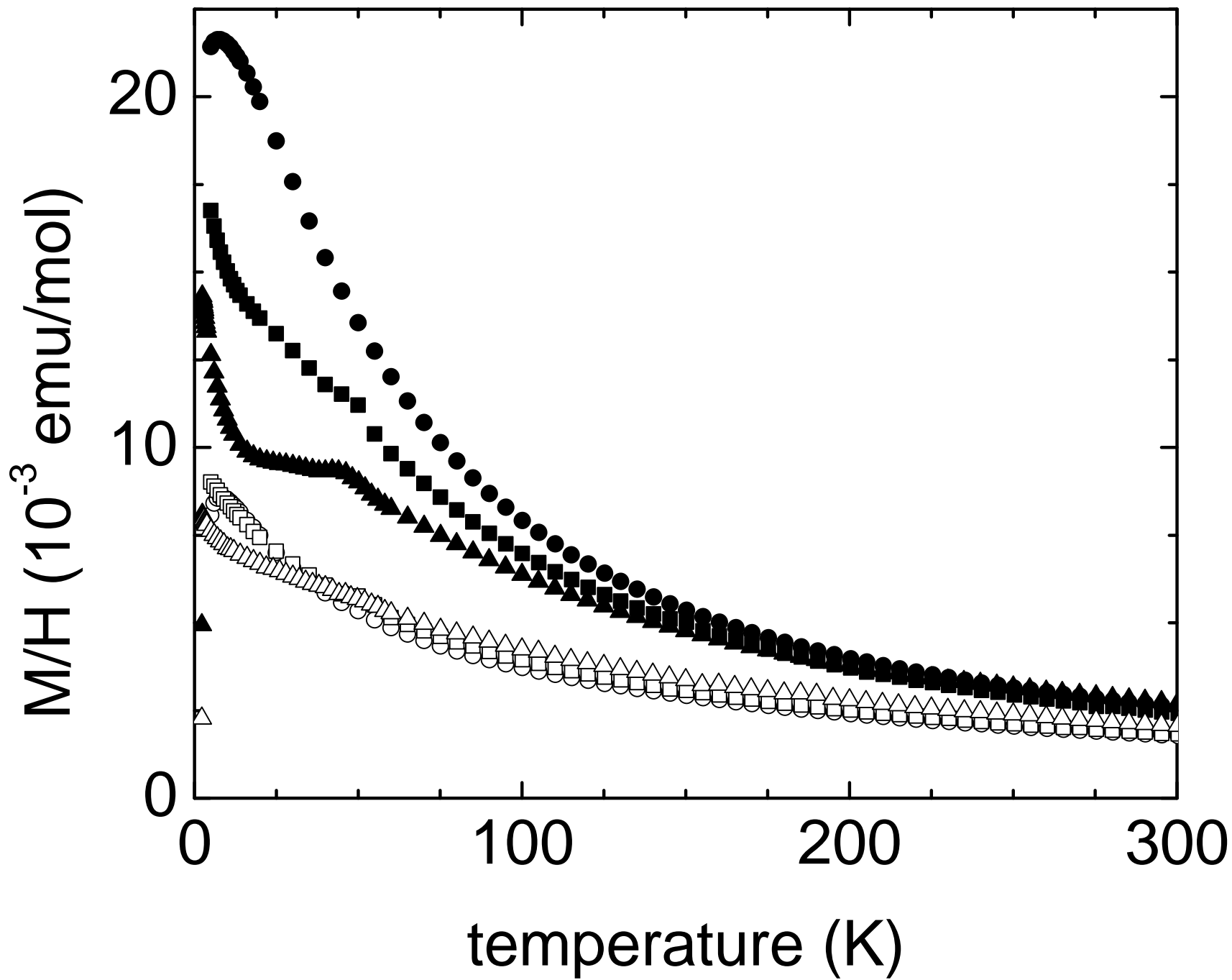

Fig. 4 Petrovic, *et al.*